\documentstyle[aps,prb,multicol,epsf]{revtex}
\def\({\begin{equation}}
\def\){\end{equation}}
\begin{document}                
\title{Magnetic fluctuations in 2D metals close to the Stoner instability}
\author{B. N. Narozhny and I. L. Aleiner}
\address{Department of Physics and Astronomy,
SUNY at Stony Brook, Stony Brook, NY 11794}
\author{A.I. Larkin}
\address{Theoretical Physics Institute, University of Minnesota, Minneapolis, 
MN 55455}
\address{L.D. Landau Institute for Theoretical Physics, 117940 Moscow, Russia}

\maketitle
\begin{abstract}
We consider the effect of potential disorder on magnetic properties of a 
two-dimensional metallic system (with conductance $g\gg 1$) when interaction
in the triplet channel is so strong that the system is close to the threshold
of the Stoner instability. We show, that under these conditions there is an
exponentially small probability for the system to form local spin droplets 
which are local regions with non zero spin density. Using a non-local version
of the optimal fluctuation method we find analytically the probability 
distribution and the typical spin of a local spin droplet (LSD). In 
particular, we show that both the probability to form a LSD and its typical 
spin are independent of the size of the droplet (within the exponential 
accuracy). The LSDs manifest themselves in temperature dependence of 
observable quantities. We show, that below certain cross-over temperature the 
paramagnetic susceptibility acquires the Curie-like temperature dependence, 
while the dephasing time (extracted from magneto-resistance measurements) 
saturates.
\end{abstract}
\pacs{PACS numbers: }

\begin{multicols}{2}

\section{Introduction}
The itinerant ferromagnetism in metals has been the subject of extensive
theoretical and experimental studies. In clean systems the ferromagnetic
instability is described by the Stoner criterion \cite{sto,pom}, which defines 
the critical value of the spin-exchange interaction constant (at which the 
system becomes unstable with respect to ferromagnetic ordering). The value 
of the constant depends on the material. In most simple metals it is small (so
that the electronic system exhibits paramagnetic response), while in palladium 
it approaches the critical value \cite{pal}. Theoretically, the Stoner 
criterion is most easily obtained within the framework of the Hubbard model 
\cite{hub} or in Landau Fermi liquid theory \cite{pom,aaa}.

In two dimensions disorder tends to localize the electronic system 
\cite{lee,alt}. However, if either the sample size $L$ or the dephasing length 
$L_\varphi$ is smaller than the localization length, the sample can still be 
considered as metallic. However, the interaction constants are renormalized 
from their clean values. In particular, the spin-exchange interaction constant 
was shown to flow towards strong coupling \cite{fin}. It is possible therefore
to have a sample with the renormalized constant approaching the critical 
value while at the same time far from the Anderson localization. In this case 
a mean-field structure of the ferromagnetically ordered ground state was 
conjectured recently in Ref.~\onlinecite{and}. 
   
An important issue in the physics of disordered systems is the role of 
mesoscopic fluctuations. Indeed, the Stoner criterion in its usual form 
reflects the tendency of the system to acquire {\it uniform} non zero 
magnetization (if the interaction constant happens to reach the critical 
value). However, for each realization of disorder the spin exchange 
interaction is non-local and random. The effective interaction constant 
entering the Stoner criterion is essentially the interaction kernel averaged
over the whole system. At the same time, averaging over some small part of the 
system might produce a value of the effective interaction constant different 
from the system-wide average. In particular, some rare impurity configurations 
can lead to the locally averaged interaction constant in some region to satisfy
the Stoner criterion, while the system-wide average does not. This would mean 
the appearance of non zero spin polarization in such regions. The similar 
effect in finite-size mesoscopic systems was considered in 
Ref.~\onlinecite{kur}.   
   
In this paper we investigate the plausibility of such a scenario and its effect
on magnetic and transport properties of the system. We consider a good metal 
(characterized by a large dimensionless conductance 
${g = 2\pi\hbar/e^2 R_\Box\gg 1}$) in 2D close to the 
instability, but still in the paramagnetic phase (at the mean field level), so 
that the naive mean-field value of the total magnetization of the system is 
zero. We describe the spin
exchange interaction by the effective averaged constant $F_0$ with all the 
renormalizations already included (as shown in Ref.~\onlinecite{fin}) and by 
the random, sample specific non-local susceptibility, which, when averaged
over the area of a small region, gives the local effective interaction 
constant. Using the ``optimal fluctuation'' method \cite{ztl} we find that 
there is exponentially small but non zero probability to find such region with
non zero spin polarization which we call a local spin droplet (LSD). With the
exponential accuracy this 
probability does not depend on the size of a LSD, therefore droplets of all 
sizes (up to the size of the order of the thermal length 
$L_T=\sqrt{\hbar D/T}$, ($D$ is the diffusion constant and $T$ is the 
temperature) can appear. The total spin of a LSD is also independent of its 
size. The effective interaction between LSDs is determined by the correlations 
of the same non-local susceptibility. Since LSDs are extended objects, the 
correlation functions which determine the effective interaction constant have 
to be averaged over the area of both interacting LSDs. The oscillating parts 
of the correlation functions (which usually lead to the RKKY \cite{rkk} 
interaction) do not survive this averaging. Instead, we find that the average
value of the effective interaction is ferromagnetic and decays with the 
distance between LSDs only as a power law. However, at large distances 
fluctuations of the effective interaction constant exceed the average and the
sign of the interaction becomes random.

The contribution of LSDs to physical observables manifests itself at low 
temperatures. Indeed, as any system of weakly interacting moments at 
temperatures, higher than the point of magnetic ordering, the system of
LSDs exhibit the Curie-like susceptibility. We show, that at not so low 
temperatures, this contribution exceeds the (temperature-independent) Pauli
susceptibility of the electron system. Likewise, the LSDs contribute to the
dephasing time $\tau_\varphi$ (extracted from magnetoresistance measurements, 
see Ref.~\onlinecite{alt}). Only, while usually \cite{alt,us} the 
dephasing time in two dimensional electron systems behaves like
$\tau_\varphi^{-1}\sim T$, the contribution of LSDs is temperature independent.
Thus at temperatures lower than certain cross-over temperature the dephasing 
time saturates. The cross-over temperature is roughly the same for both 
quantities. When $T\rightarrow 0$, interactions of LSDs with each other or with
itinerant electrons should lead either to the screening of the local spins or 
to forming of some spin glass state (due to the randomness of the interaction).
We do not consider such regime in this paper.

The paper is organized as follows. In Section~\ref{stoner} we review the basic
physical description of the Stoner instability, establish notations and, in
subsection~\ref{mf} we outline the effect of mesoscopic fluctuations. Next,
in Section~\ref{qual} we discuss the formation of LSDs qualitatively. The
subsection~\ref{optimal} is devoted to the calculation of the probability to 
form a LSD. Then, in subsection~\ref{spdis} we find the distribution of the
total spin of the LSD by means of the ``optimal fluctuation'' method 
\cite{ztl}, which we adapt to the non-linear problem. Section~\ref{inter} 
describes the interaction of LSDs, and in Section~\ref{phys} we discuss the
contribution of LSDs to physical observables. Our results are summarized in
Conclusions. Some mathematical details of the non-local susceptibility 
correlations are relegated to the Appendix.

\section{Stoner Instability}
\label{stoner}

The purpose of this Section is to recall the basic ideas leading to the Stoner 
criterion (subsections~\ref{rps} and \ref{gl}) and to contrast the situation 
in clean systems to that in disordered metals. In subsection~\ref{mf} we 
demonstrate the role of mesoscopic fluctuations and discuss how
they can lead to formation of LSDs. For further details on Stoner 
ferromagnetism the reader is referred to the standard textbooks 
Ref.~\onlinecite{hub,aaa}.

\subsection{Renormalized paramagnetic susceptibility}
\label{rps}

The paramagnetic response of a system of non-interacting electrons is 
described by the Pauli susceptibility which depends only on the electronic 
density of states at the Fermi level $\nu$. This can be seen from the 
following observation. The electron energy $\epsilon$ enters all thermodynamic 
functions in combination $\epsilon - \mu$ with the chemical potential $\mu$. 
The interaction energy of the electron spin $\vec s$ with the external magnetic
field $\vec h$ (which is proportional to $-\vec s \vec h$) can thus be 
considered as a shift of the chemical potential. Since it is proportional to 
$\vec s$, the number of electrons which spin is aligned with $\vec h$ exceeds 
the number of electrons with the opposite spin, resulting in the total 
magnetization proportional to $\nu\vec h$.

Although quite general, the above argument relies on the fact that weak 
magnetic field does not change the energy spectrum of the electron system.
Taking into account the electron-electron interactions, however, changes the 
distribution function of electrons and thus the electronic energy spectrum.
As a result, some physical quantities become renormalized from their bare
values. In particular, the paramagnetic susceptibility is 
renormalized by the
the spin exchange interaction \cite{aaa}(we choose the units with Bohr 
magneton equal to unity) 

\begin{equation}
\chi = {{\nu}\over{1+F_0}}.
\label{pauli}
\end{equation}
 
\noindent
Here the parameter $F_0$ is the effective dimensionless coupling constant of 
magnetic interaction between electron spins. Within the phenomenology of the
Landau Fermi-liquid theory\cite{aaa} it can be obtained by averaging the 
spin-exchange 
part of the Landau function over the Fermi surface. In the case when $F_0<0$, 
the interaction tends to align the electrons spins, competing with the Pauli 
exclusion principle. If the interaction is strong enough, the gain in the 
magnetic energy exceeds the kinetic energy cost needed to realign the spins 
and the ground state of the system changes to the one with non zero total spin 
- it becomes ferromagnetic. According to Eq.~(\ref{pauli}) the instability 
occurs when $F_0 = -1$, which is known as the Stoner criterion\cite{sto,pom}. 

More formally, the full susceptibility tensor is given by the commutator of 
the spin density operators $\hat\sigma_\alpha(x, t)$ (hereafter $x_i$ denotes 
the two-component coordinate vector) 

\begin{eqnarray}
\chi_{\alpha\beta}(x_1, && x_2; t_1-t_2) \nonumber\\
&&
\nonumber\\
&&
= i\theta(t_1-t_2)
\langle[\hat\sigma_\alpha(x_1, t_1), \hat\sigma_\beta(x_2, t_2)]\rangle,
\label{chid}
\end{eqnarray}

\noindent
where $\alpha, \beta = x, y, z$. In the 
paramagnetic state of an isotropic system the tensor $\chi_{\alpha\beta}$ is 
diagonal and isotropic and can be expressed in terms of the transverse 
susceptibility $\chi$, 

\begin{eqnarray*}
\chi_{\alpha\beta} = 2 \chi \delta_{\alpha\beta},
\end{eqnarray*}

\noindent
which is determined in terms of the commutator of the spin
raising and lowering operators $\hat\sigma^+$ and $\hat\sigma^-$ similarly to 
Eq.~(\ref{chid}). 

The transverse susceptibility can be evaluated in the generalized Hartree-Fock 
approximation \cite{hub}, which amounts to summation of ladder diagrams in the 
particle-hole channel. In the Galilean invariant system the susceptibility
depends only on the coordinate difference and in the momentum representation
is given by \cite{hub}

\begin{equation}
\chi(q, \omega) = {{\Pi(q, \omega)}\over{1+U\Pi(q, \omega)}}.
\label{chis}
\end{equation}

\noindent
Here $\Pi(q, \omega)$ is the electron polarization operator, which represents
the susceptibility of the non-interacting electron gas. In the limit
$\omega=0$ and $q=0$ it gives the Pauli susceptibility [since 
$\Pi(q=0, \omega=0) = \nu$]. The parameter $U$ is the spin exchange 
coupling constant. In the context of the Hubbard model it appears as a 
phenomenological parameter of the Hamiltonian. In the microscopic Fermi-liquid
theory it is the spin exchange part of the vertex function $\Gamma^\omega$ 
averaged over the Fermi surface (compare with the scattering amplitude 
$\Gamma_2$ in Ref.~\onlinecite{fin}). The instability in the ground state of 
the system corresponds to the singularity in the static limit of the response 
function $\chi(q, \omega=0)$. The instability criterion is thus
$U\Pi(q, \omega=0) = -1$. At $q=0$ this corresponds to a tendency of the system
to acquire spontaneously a uniform (or ferromagnetic) spin density. The 
criterion for this instability 
 
\begin{equation}
U\nu = -1.
\label{crit}
\end{equation}

\noindent
corresponds to the Stoner criterion if one identifies the Landau parameter
$F_0$ with the spin exchange coupling constant.

\subsection{Magnetization energy}
\label{gl}

The transition to the ferromagnetic state can also be described similarly to
the Landau description of the phase transitions with the induced
spin density as the order parameter. Close to the instability point, the 
thermodynamic potential $\Omega$ of the system can be expanded in powers of
the spin density $\vec\sigma$

\begin{equation}
\Omega = \int \frac{dV}{2\nu} \left [
(1+F_0)\vec\sigma^2 + a^2 |\nabla \vec\sigma|^2 
+\frac{1}{2}B (\vec\sigma^2)^2 + ... \right ],
\label{tp}
\end{equation}

\noindent
where $B>0$. The value of $\vec\sigma$ of any particular state of the system 
(as described by the constants $F_0$, $a$, and $B$) can be found by minimizing 
the thermodynamic potential $\Omega$. In the limit $q=0$ (uniform spin 
density) the minimum condition is

\begin{equation}
(1 + F_0 + B \vec\sigma^2)\vec\sigma = 0.
\label{eneq}
\end{equation}

\noindent
Here we have neglected the gradient term relative to the linear term, since 
the coefficient \cite{lam} $a\sim k_F^{-2}$ and under our assumptions, $g\gg 1$
and $g(1+F_0)\gg 1$ (the latter assumption will be elaborated on in 
Section~\ref{qual}). Therefore the linear term dominates, $1+F_0 > k^2/k_F^2$
even for large momenta $k\sim 1/l$ ($l$ is the mean free path). 
As usual, in order to have a non-trivial solution ${\sigma\ne 0}$, one should
have $1+F_0 < 0$, which is again the Stoner criterion. In the ferromagnetic 
phase the solution to Eq.~(\ref{eneq}) gives the total value of the induced
spin density ${\sigma^2 = -(1+F_0)/B}$.

\subsection{Mesoscopic fluctuations}
\label{mf} 

In the presence of disorder the local spin density $\sigma(x)$ depends on the
particular impurity distribution and before averaging can be taken as random.
In the paramagnetic susceptibility~(\ref{chis}) the coupling $U$ is 
determined by small distances and thus does not depend on disorder. On the 
contrary, the polarization operator $\Pi(x_1, x_2, \omega)$ includes large 
distances (since we are interested in the limit $q=0$) and is thus strongly 
affected by disorder. Consequently, the paramagnetic susceptibility is random. 
Moreover, it depends on both coordinates since translational invariance is lost
and it is also non-local. 
However, if disorder is not too strong, then the 
non-local, random part of the susceptibility can be separated from the uniform
term, which is independent of disorder and represents the susceptibility of 
the clean system. Preserving the form of Eq.~(\ref{pauli}), the susceptibility 
of the disordered system can be found as a solution to the integral equation

\begin{eqnarray}
\int d^2x_2 [(1+F_0)\delta(x_1 - x_2) &&+ F_1(x_1, x_2)] \chi(x_2, x_3) 
\nonumber\\
&&
\nonumber\\
&&
= \nu \delta(x_2 - x_3),
\label{chin}
\end{eqnarray}
 
\noindent
where $F_1(x_1, x_2)$ is a random quantity with zero mean. 

Similarly, the thermodynamic potential (\ref{tp}) becomes 

\begin{eqnarray}
\Omega \; && =
\frac{1}{2\nu}\int d^2x(1+F_0)\vec\sigma^2(x) \nonumber\\
&&
\nonumber\\
&&
+ \frac{1}{2\nu}\int d^2x_1 d^2x_2F_1(x_1, x_2)
\Big(\vec\sigma(x_1)\vec\sigma(x_2)\Big) \\
\label{tpnl}
&&
\nonumber\\
&&
+ \frac{1}{4\nu}\int\prod\limits_{i=1}^4 d^2x_i B\left[\{x_j\}\right]
\Big(\vec\sigma(x_1)\vec\sigma(x_2)\Big)
\Big(\vec\sigma(x_3)\vec\sigma(x_4)\Big).\nonumber
\end{eqnarray}

\noindent
Again, we are interested in the limit $q\rightarrow 0$ (neglecting the 
gradient term; see the previous subsection and an estimate below). In the 
second order term we have neglected the possibility of the presence of 
spin-orbit coupling. While the spin-orbit interaction can be taken into 
account, its presence does not affect our main results (see 
Section~\ref{inter} for discussion). In addition to the fourth-order term 
written in Eq.~(\ref{tpnl}) there is a term with different spin structure, 
namely $\left [ \vec\sigma(x_1)\times\vec\sigma(x_2)\right ]
\left [ \vec\sigma(x_3)\times\vec\sigma(x_4)\right ]$ (see Appendix for 
details). In what follows we assume the simplest spin structure for a
LSD $\vec\sigma = (0, 0, \sigma)$. In this case the additional cross
product term vanishes and we can treat the spin density as a scalar. 

Similarly to Eq.~(\ref{eneq}) the minimum of $\Omega$ can be found from the 
(now non-local) integral equation

\begin{eqnarray}
(1&&+F_0)\sigma(x_1) 
+ \int d^2x_2F_1(x_1, x_2)\sigma(x_2) \nonumber\\
&&
\nonumber\\
&& + 
\int\prod\limits_{i=2}^4 d^2x_i B\left[\{x_j\}\right]
\sigma(x_2)\sigma(x_3)\sigma(x_4) = 0.
\label{eq}
\end{eqnarray}

\noindent
The coefficient $B\left[\{x_j\}\right]$ in the thermodynamic potential 
Eq.~(\ref{tpnl}) is also random. In this paper we take both $F_1$ and $B$ to 
be Gaussian random matrices, with the distribution, which in compactified 
notation is given by

\begin{eqnarray}
w[F_1, B] \propto \exp\left[ -\int 
\pmatrix{
F_1 ; & B \cr
 }
\hat{\cal K}^{-1}
\pmatrix{
F_1 \cr
B 
 }
\right]
\label{dist}
\end{eqnarray}

\noindent
The Gaussian approximation is valid while the expression in the exponent
does not exceed the dimensionless conductance $g$, where the log-normal tail
appears \cite{ler}.
The integration is over all the variables of $F_1$ and $B$. The distribution 
$w[F_1, B]$ should be understood in the operator sense and will be used to 
evaluate the functional integrals below. The weight operator
$\hat {\cal K}$ is constructed from the correlators

\begin{eqnarray}
\hat{\cal K} = 
\pmatrix{
\langle F_1 F_1 \rangle & \langle F_1  B \rangle \cr
\langle B   F_1 \rangle & \langle B    B \rangle \cr
 },
\label{corr}
\end{eqnarray}

\noindent
which are discussed in detail in Appendix, where we give their explicit form. 

For a given realization of disorder, the equation~(\ref{eq}) might allow 
for some non-trivial solution $\sigma^{(0)}(x)$. To estimate the value of 
the total spin corresponding to such a solution, we write the spin density
as

\begin{equation}
\sigma^{(0)}(x) = \sigma_0 \psi(x),
\label{sep}
\end{equation}

\noindent
where $\psi(x)$ is normalized to unity, therefore both $\psi(x)$
and $\sigma_0$ have dimension of inverse length. The total value of the 
spin is determined by 

\begin{equation}
S = \sigma_0 \int d^2x \psi(x),
\label{spin}
\end{equation}

\noindent
while $\sigma_0$ can be found from Eq.~(\ref{eq}) (in the case when it allows 
a non-trivial solution) in the form

\begin{equation}
\sigma_0^2 = - \frac{1+F_0+F_1^{(0)}}{B^{(0)}}.
\label{s0}
\end{equation}

\noindent
Here the constants $F_1^{(0)}$ and $B^{(0)}$ are the ``matrix elements'' 
of the non-local operators $F_1(x_1, x_2)$ and $B\left[\{x_j\}\right]$
(if $\psi(x)$ is interpreted as a ``wave function'')

\begin{mathletters}
\begin{equation}
F_1^{(0)} = 
\int d^2x_1 d^2x_2F_1(x_1, x_2)\psi(x_1)\psi(x_2)
\label{mef}
\end{equation}
\begin{equation}
B^{(0)} = \int\prod\limits_{i=1}^4 d^2x_i B\left[\{x_j\}\right]
\prod\limits_{k=1}^4\psi(x_k).
\end{equation}
\label{me}
\end{mathletters}

Consider now a metal close to the Stoner instability, so that $0<1+F_0\ll 1$. 
While, as follows from Eq.~(\ref{eneq}), the averaged, uniform spin density 
is zero, solutions~(\ref{s0}) to the non-local equation~(\ref{eq}) might  
exhibit non zero spin density in some rare regions, where due to a particular
configuration of impurities the non-local part of the susceptibility $F_1$ is 
negative. If there are several regions with non zero total spin, then the
spin-spin or magnetic interaction between them will contribute to the ground 
state energy of the system and correspondingly to the magnetic susceptibility.
If such interaction favors some kind of ordering of the spins, then the
appearance of these regions can change the magnetic response of the system, 
in other words change the ground state. 

Fluctuation effects in systems close to a phase transition have been studied
extensively (see, for instance, Ref.~\onlinecite{lam,lev,spz,spi}). In 
particular, the picture of smearing the transition point by formation of 
fluctuation regions with non-zero value of an order parameter was considered 
in Ref.~\onlinecite{lev}. In order to determine a value of the order
parameter in a fluctuation region a solution of the non-linear Ginsburg-Landau 
equation in the presence of disorder was needed. Two issues make this case 
different from ours. First, in Ref.~\onlinecite{lev} the fourth-order term $B$
was not random.  Second, the fluctuations of the order parameter were assumed
to be local. Therefore, the white noise approximation (i.e. approximating the
correlatino functions~(\ref{corr}) by the delta-functions) for the disorder 
was appropriate. As a result, the fluctuation regions did not interact and the
percolation scenario of the phase transition was needed. In our case the 
non-locality of fluctuations (expressed in terms of the correlation 
functions~(\ref{corr}) leads to interaction between LSDs which results in a
change of behavior of observable quantities as discussed in Section~\ref{phys}.

\section{Qualitative discussion}
\label{qual}

In the previous subsection~\ref{mf} we indicated how LSDs - local regions
with non zero spin polarization - could appear in a metal close to the Stoner 
instability due to fluctuations in impurity distribution. Here we estimate
qualitatively the probability to find a LSD and the value of its total spin.
     
Treating a LSD as an open region of the size $R$ we can characterize it by
the Thouless energy $E_T = D/R^2$ (where $D$ is the diffusion constant). The
inverse of the Thouless energy is the ``escape time'' $\tau_{esc} = E_T^{-1}$,
which is 
the time it takes for the diffusing particle to leave the LSD. This time scale 
serves as the infrared cut-off for the correlation function, which describes
the mesoscopic fluctuations of the density of states (DoS) \cite{shk}

\begin{eqnarray}
\langle \rho (\epsilon ) \rho (\epsilon + \omega ) \rangle &&\simeq
{\mathrm Re} \int \frac{R^2d^2Q}{(-i\omega + DQ^2 + \tau_{esc}^{-1})^2}
\nonumber\\
&&
\nonumber\\
&&
= \frac{\pi}{E_T} {\mathrm Re} \frac{1}{-i\omega+\tau_{esc}^{-1}}.
\label{cdos}
\end{eqnarray}

\noindent 
The magnetic energy of the LSD, written in terms of its total spin $S$ is

\begin{equation}
E(S) = \delta_1 (1+F_0) S^2 + \delta E(S),
\end{equation}

\noindent
where $\delta_1$ is the mean level spacing and $\delta E(S)$ denotes 
contribution of all non-linear terms in Eq.~(\ref{tpnl}). All the terms 
contributing to $\delta E(S)$ are random and can be expressed in terms of the 
fluctuating (random) DoS

\begin{eqnarray}
\delta E(S) = \int\limits_0^{\delta_1 S} ds_1 \int\limits_0^{s_1} ds_2
\left[ \rho(s_2) + \rho(-s_2) \right].
\end{eqnarray}

\noindent
The averaged $\delta E(S)$ equal to zero, but the average of its square
$\langle (\delta E(S))^2 \rangle$ is not and it is determined by the 
correlator~(\ref{cdos})
 
\begin{eqnarray}
\langle (\delta E(S))^2 \rangle \simeq
\left\{
\matrix{
\frac{\delta_1^2 S^4}{g^2} ,  & S \ll g ;\cr
\frac{\delta_1^2 S^3}{g} , & S \gg g .\cr
}
\right.  
\label{fs}
\end{eqnarray}

In this paper we restrict ourselves to consideration of metals close to the 
instability point, where the overall spin of the LSD is small $S \ll g$.
In this case we can treat $\delta E(S)$ as a Gaussian random quantity.
Moreover, we can expand it in powers of $S/g$ so that the magnetic energy of
the LSD becomes

\begin{equation}
E(S) = \delta_1 (1+F_0) S^2 + 
\zeta \left[ \frac{S^2}{g} - \frac{S^4}{g^3} \right],
\label{es}
\end{equation}

\noindent
where $\zeta$ is a random Gaussian variable with the distribution

\begin{eqnarray}
P(\zeta) \propto \exp\left( -\zeta^2\right).
\label{dg}
\end{eqnarray}

\noindent
Note that there is only one random quantity $\zeta$ in Eq.~(\ref{dg}). The 
equation~(\ref{es}) is valid only for $S\ll g$. That is why energy minima
at$\zeta > 0$ are spurious and should not be considered.

To find the distribution of the spin value, we need to minimize the energy
$E(S)$. Differentiating Eq.~(\ref{es}) we obtain the equation

\begin{equation}
\left ( 1 + F_0 + \frac{\zeta}{g} \right) S - \zeta\frac{S^3}{g^3} = 0.
\label{eqg}
\end{equation}

\noindent
This equation allows for non-trivial solutions $S>0$ when $\zeta<-g(1+F_0)$.
As was noted above, we are working in the regime where $g(1+F_0)\gg 1$, so that
the fluctuation that creates the LSD is rare indeed. To determine the 
distribution of spin values, we solve the equation~(\ref{eqg}) for $\zeta$
and substitute in the Gaussian distribution Eq.~(\ref{dg}). As a result, we
estimate the distribution (up to numerical coefficients)

\begin{equation}
P(S) = \exp\left(-\frac{g^2(1+F_0)^2}{(1-\frac{S^2}{g^2})^2}\right).
\label{r11}
\end{equation}

\noindent
This distribution is only valid when $g(1+F_0)\gg 1$ and $S\ll g$, therefore
we can expand the denominator in the exponent without loss of accuracy

\begin{equation}
P(S) = \exp\left [- g^2(1+F_0)^2 -  (1+F_0)^2 S^2 \right].
\label{r12}
\end{equation}

Integrating over the spin $S$ we estimate the probability to find a LSD 

\begin{equation}
{\cal P} \propto \exp\left (- g^2(1+F_0)^2 \right).
\label{r10}
\end{equation}

\noindent
It is determined by the first term in the exponent Eq.~(\ref{r12}). The second 
term determines the typical value of the spin of the LSD

\begin{equation}
S\simeq \frac{1}{1+F_0}.
\label{r13}
\end{equation}

\noindent
Remarkably, this value and the probability Eq.~(\ref{r10}) do not depend on 
the size $R$ of the LSD, which is the main qualitative result of this Section. 
Note that it is similar to the result for zero-dimensional grains 
\cite{kur}. Also the spin value is independent of the dimensionless 
conductance $g$. These facts determine the contribution of LSDs to the 
physical observables considered below in Section~\ref{phys}.  

The applicability of the consideration of this Section and quantitative 
results of Section~\ref{lsds} is limited by two requirements. First, the 
probability Eq.~(\ref{r11}) must be  exponentially small,
so that $g(1+F_0)\gg 1$. Second, the Gaussian approximation Eq.~(\ref{dg}) 
[similarly to the distribution Eq.~(\ref{dist})] is valid while the 
expression in the exponent is smaller than the dimensionless conductance of
the system \cite{ler}, so that in Eq.~(\ref{r11}) $g^2(1+F_0)^2 < g$. 
Combining the two limits, we obtain the region of applicability of the results 
Eqs.~(\ref{r11}) - (\ref{r13}) as 

\begin{equation}
1/g \ll 1+F_0 < 1/\sqrt{g}.
\label{app}
\end{equation}

\section{Local spin droplets}
\label{lsds}

In this section we calculate the probability to find a local region with non 
zero spin, which we call a local spin droplet (LSD) and the value of the total
spin of the LSD. The spin and the spatial profile of the LSD can be found from
the non-linear equation~(\ref{eq}). In subsection~\ref{optimal} we show 
that to the exponential accuracy the probability to find the LSD 
Eq.~(\ref{r10}) is captured by the linear part of Eq.~(\ref{eq}), while the
non-linear term fixes the spin value, as shown in subsection~\ref{spdis}.

\subsection{Probability to form a LSD}
\label{optimal}

The calculation of the 
probability to find a rare fluctuation leading to formation of a LSD can be 
performed along the lines of the argument used in Ref.~\onlinecite{ztl} to 
calculate the exponentially small tail in the density of states (DoS) of a 
particle in a random potential \cite{ztl}. In the quantum mechanical problem, 
considered in Ref.~\onlinecite{ztl}, one looks for such fluctuation of the 
random potential that creates a low energy bound state, thus leading to non 
zero DoS at that energy. The probability to form the bound state is determined 
by the distribution of the matrix elements of the random potential. While 
being exponentially small, the probability should be maximized by choosing 
the ``optimal'' fluctuation of the potential.  
 
To gain some intuition about how the optimal fluctuation method can be
applied to the problem at hand, in this section we consider the linear part of
the equation (\ref{eq}), disregarding for a moment the higher order $B$
term. Such an approach can be justified by observing that close to the 
instability the non-linear term in Eq.~(\ref{eq}) is small compared to the
linear ones, since the induced spin density on average is equal to zero. The
non-linear term stabilizes a non-trivial solution and fixes its amplitude, 
while the existence of such a solution can be uncovered at the level of the 
linear problem. Thus the linear equation captures the main contribution to the 
probability and at the same time demonstrates the similarity of our problem to 
the problem of tails in the DoS as well as the peculiar differences.

We can write the linear equation in the operator form 

\begin{equation}
\hat F_1 \psi(x) = E \psi(x).
\label{leq}
\end{equation}

\noindent
We write $\psi(x)$ instead of $\sigma(x)$ to stress the point that the linear
equation does not allow us to determine the value of the spin but only the
spatial profile of the LSD. Therefore, Eq.~(\ref{leq}) is simply the 
eigenproblem for the operator $F_1$ and as such does not fix the normalization
of eigenfunctions $\psi(x)$, which we are free to normalize to unity for
convenience. Since the eigenvalue problem Eq.~(\ref{leq}) is similar to the 
quantum mechanical problem of Ref.~\onlinecite{ztl}, we can adopt the 
language of the Schr\"odinger equation,  with the (now integral) operator 
$\hat F_1$ playing the role of the ``Hamiltonian'', $E$ the ``energy'' and
$\psi(x)$ the wave-function. 

For some particular realizations of the random potential in Eq.~(\ref{leq})
there is a low energy bound state with the energy $E_0[\{F_1\}] = F_1^{(0)}$
[given by Eq.~(\ref{mef})], resulting in
non zero DoS at this energy. For energies close to $E_0[\{F_1\}]$ only the 
bound state contributes to the DoS, which before averaging over disorder is 
given by the single delta-function

\begin{equation}
\rho(E) = \delta (E - E_0[\{F_1\}]).
\label{del}
\end{equation}

\noindent
Averaging this DoS over disorder takes into account contributions of all
possible realizations of the random potential leading to such bound states
and thus results in an exponentially small but smooth function of energy $E$.
This function is proportional to the probability to find the bound state at
energy $E$. In particular, for the special value $E=-(1+F_0)$, it would 
give the probability to find the non-trivial solution to the linear part of
Eq.~(\ref{eq}) (or to find the LSD).

The random quantity in the linear problem Eq.~(\ref{leq}) is $F_1$ itself.
Its distribution is obtained from Eq.~(\ref{corr}). The averaged probability 
is then

\begin{mathletters}
\begin{eqnarray}
\langle \rho (E) \rangle = &&\int {\cal D}[F_1] \delta (E - E_0[\{F_1\}])
\nonumber\\
&&
\nonumber\\
&&
\times\exp\left[ -\int  d^2x_1 d^2x_2 d^2y_1 d^2y_2 \; A \right],
\end{eqnarray}
\begin{eqnarray}
A = F_1(x_1, x_2) 
K_{FF}^{-1}\left[\{x_j\}, \{y_j\}\right] F_1(y_1, y_2),
\label{aps}
\end{eqnarray}
\label{ap}
\end{mathletters}

\noindent
where $K_{FF}^{-1}\left[\{x_j\}, \{y_j\}\right]$ is the inverse of the 
correlator Eq.~(\ref{ff}) i.e.

\begin{eqnarray*}
\int  d^2y_1 d^2y_2 K_{FF}^{-1}\left[\{x_j\}, \{y_j\}\right] &&
K_{FF}\left[\{y_j\}, \{z_j\}\right] = \nonumber\\
&&
\nonumber\\
&&
\delta(x_1-z_1)\delta(x_2-z_2).
\end{eqnarray*}

In the ``optimal fluctuation'' approach one has to evaluate the integral 
Eq.~(\ref{ap}) in the saddle point approximation. To find the saddle point 
one has to minimize the exponent $A$ of the Gaussian probability with respect 
to all functions $F_1(x_1, x_2)$ subjected to the condition 

\begin{equation}
E=E_0[\{F_1\}],
\label{co}
\end{equation}

\noindent
represented by the delta-function in Eq.~(\ref{ap}). This involves solving 
the equations

\begin{eqnarray*}
\frac{\delta}{\delta F_1(x_1, x_2)} \left [ A + \lambda E_0[\{F_1\}] \right ]
=0,
\end{eqnarray*}

\noindent
where $\lambda$ is the Lagrange multiplier to be found from the condition 
Eq.~(\ref{co}).

The saddle point solution for $F_1$ represents the optimal 
fluctuation of the random potential, 

\begin{eqnarray}
\bar F_1(&&x_1, x_2) = 
\nonumber\\
&&
\nonumber\\
&&
\lambda \int  d^2z_1 d^2z_2 
K_{FF}\left[\{x_j\}, \{z_j\}\right]
\psi(z_1) \psi(z_2),
\label{ofl}
\end{eqnarray}

\noindent
given in terms of the eigenfunction $\psi(x)$ corresponding to the eigenstate 
$E_0$. Substituting Eq.~(\ref{ofl}) into Eq.~(\ref{leq}) we obtain

\begin{eqnarray}
E\psi(x_1) = \lambda\int &&d^2x_2 d^2z_1 d^2z_2 \psi(x_2)
\nonumber\\
&&
\nonumber\\
&&
\times K_{FF}\left[\{x_j\}, \{z_j\}\right]
\psi(z_1) \psi(z_2),
\label{nleq}
\end{eqnarray}

\noindent
which, together with the normalization condition $\int d^2x |\psi(x)|^2 = 1$, 
constitute the analogue of the non-linear Schr\"odinger equation of 
Ref.~\onlinecite{ztl}.

Both the saddle point value of $\lambda$ and $\psi(x)$ should be found from 
the non-linear equation (\ref{nleq}) and the normalization condition. The 
averaged DoS is given by the Gaussian probability in Eq.~(\ref{ap}), evaluated 
at the saddle point,

\begin{eqnarray}
\langle \rho (E) \rangle \sim \exp \left( - \frac{E^2}{2I_{FF}} \right),
\label{p}
\end{eqnarray}

\noindent
where 

\begin{eqnarray}
I_{FF} = && \int  d^2z_1 d^2z_2 d^2x_1 d^2x_2 
\nonumber\\
&&
\nonumber\\
&&
\times\psi(x_1) \psi(x_2)
K_{FF}\left[\{x_j\}, \{z_j\}\right]
\psi(z_1) \psi(z_2).
\label{int}
\end{eqnarray}

\noindent
The integral $I_{FF}$ is dimensionless (independent of any length scale)
since we require the eigenfunction $\psi(x)$ to be normalized, so that
$\psi(x) \sim 1/R$, where $R$ is roughly the size of the LSD. As we are 
discussing the single LSD, all four of the eigenfunctions in Eq.~(\ref{int}) 
are centered around approximately the same point, therefore $R$ is the only
scale in Eq.~(\ref{int}). In this case the dependence of the correlator 
$K_{FF}$ on $R$ is given by Eq.~(\ref{dk}), $K_{FF}\propto g^{-2}R^{-4}$. Thus
the integral Eq.~(\ref{int}) and the probability Eq.~(\ref{p}) are independent
of the size of the LSD $R$. 

The probability to find the LSD is given by Eq.~(\ref{p}), evaluated at the 
point $E=-(1+F_0)$,

\begin{eqnarray}
\rho \sim e^{-\gamma g^2(1+F_0)^2},
\label{r1}
\end{eqnarray}

\noindent
where $\gamma$ is the numerical factor which is given by the dimensionless
counterpart of the integral Eq.~(\ref{int}). This result is valid while the 
number in the exponent is large, which corresponds to the lower limit of 
applicability of our consideration $1+ F_0 \gg 1/g $ (see the last paragraph
of Section~\ref{qual}). 

To determine the numerical coefficient $\gamma$ in Eq.~(\ref{r1}) we need to 
know the precise form of the eigenfunction $\psi(x)$. We could not solve the
non-linear integral equation~(\ref{nleq}) analytically and used a variational
approach. Since the kernel $K_{FF}$ in the integral Eq.~(\ref{int}) decays as 
$R^{-4}$ at large distances, the optimal $\psi(x)$ is a limited-range function 
(i.e. its normalization integral is determined on a limited interval of $x$). 
To estimate the upper limit of $\gamma$ we take the variational function of the
Gaussian form $\psi(x)=\pi^{-1/2} R^{-1} \exp(-x^2/R^2)$, substitute in 
Eq.~(\ref{int}) and evaluate the integral numerically. The resulting estimate 
is

\begin{eqnarray}
I_{FF} \approx \frac{1}{0.8 \pi^2 g^2},
\label{I1}
\end{eqnarray}

\noindent
and the numerical factor in Eq.~(\ref{r1}) is thus $\gamma \approx 3.9$.

The most important feature of the result Eq.~(\ref{r1}) is its independence on
the size $R$ of the LSD in agreement with the qualitative results of 
Section~\ref{qual}. This means that LSDs of small and large sizes can 
appear with equal (in the exponential sense) probability, given the suitable 
fluctuation of the impurity configuration. We shall return to this point below,
when we consider the interaction between LSDs.

\subsection{Spin distribution}
\label{spdis}

The argument that led us to the probability to find an LSD Eq.~(\ref{r1})
is not complete because it does not help us to determine the value of the
spin of the LSD. This follows from our consideration of the linear part
of Eq.~(\ref{eq}) only. To determine the value of the spin we must solve 
the full non-linear equation. 

The single-mode approximation to the solution of the non-linear problem was 
outlined in subsection~\ref{mf}.
The formal solution for the amplitude of the spin of a LSD is given by 
Eq.~(\ref{s0}). Again, as we did above in the case of the linear problem, we 
employ the optimal fluctuation method to find the probability to form a LSD, 
which is characterized by the total spin $S$. However, this time our task is 
simplified since the probability to form the LSD (regardless of its spin)
has already been found. We now need to find how that probability depends on
the spin value $S$. Therefore we take the function $\psi(x)$, which describes
the spatial profile of the LSD from the linear problem and focus on the 
distribution of the spin amplitude $\sigma$.

Similar to Eq.~(\ref{del}), the probability to find an LSD 
characterized by the spin density amplitude $\sigma$ is given by the 
delta-function

\begin{equation}
\rho(\sigma^2) = \delta ( \sigma^2 - \sigma_0^2[F_1, B]),
\label{dels}
\end{equation}

\noindent
which we write in terms of $\sigma^2$ for convenience. 

Averaging over disorder is performed as it was done for the linear problem
[see Eq.~(\ref{ap})]. Only now we have two random quantities, $F_1$ and $B$,
therefore we need to average with the distribution Eq.~(\ref{dist})

\begin{equation}
\langle \rho(\sigma^2) \rangle = \int {\cal D}[F_1, B] 
\delta ( \sigma^2 - \sigma_0^2) w[F_1, B],
\label{pfb}
\end{equation}

\noindent
To find the saddle point (or the optimal fluctuation) we have to minimize
the exponent

\begin{eqnarray}
A_{nl} = \lambda\Biggl( \sigma^2 + 
&& \frac{1+F_0+F_1^{(0)}}{B^{(0)}} \Biggr )
\nonumber\\
&&
\nonumber\\
&&
- \int \pmatrix{ F_1 & B \cr }\hat{\cal K}^{-1}\pmatrix{F_1 \cr B }
\label{expo}
\end{eqnarray}

\noindent
where $F_1^{(0)}$ and $B^{(0)}$ are the integrals Eq.~(\ref{me}) and $\lambda$
is again the Lagrange multiplier. The saddle point equations are given by

\begin{eqnarray}
\hat{\cal K}^{-1}\pmatrix{F_1 \cr B } = 
\frac{\lambda}{2B^{(0)}}
\pmatrix{ \psi(x_1)\psi(x_2)\cr 
\sigma^2 \psi(y_1)\psi(y_2)\psi(y_3)\psi(y_4) },
\label{spe}
\end{eqnarray}

\noindent
where the functions $\psi(x)$ appeared after differentiating the 
integrals Eq.~(\ref{me}) with respect to $F_1$ and $B$. Multiplying both the
left-hand and the right-hand sides of  Eq.~(\ref{spe}) by $\hat{\cal K}$ we 
obtain two equations with $F_1$ and $B$ in the
left-hand side only. We then multiply the first equation by two functions
$\psi(x)$ and the second by four functions $\psi(x)$ and integrate over their 
variables to obtain the algebraic equations, with integrals Eq.~(\ref{me}) as 
the unknowns

\begin{mathletters}
\begin{equation}
F_1^{(0)} = \frac{\lambda}{2B^{(0)}} 
\left( I_{FF} + \sigma^2 I_{FB} \right ),
\end{equation}
\begin{equation}
B^{(0)} = \frac{\lambda}{2B^{(0)}} 
\left( I_{FB} + \sigma^2 I_{BB} \right ),
\end{equation}
\label{spea}
\end{mathletters}

\noindent 
where $I_{FF}$ is given by Eq.~(\ref{int}) and $I_{FB}$ and $I_{BB}$
are similarly defined as

\begin{mathletters}
\begin{eqnarray}
I_{FB} = \int  d^2x_1&& d^2x_2\prod\limits_{i=1}^4 d^2y_i
\psi(x_1) \psi(x_2)
\nonumber\\
&&
\nonumber\\
&&
\times K_{FB}\left[\{x_j\}, \{y_j\}\right]
\prod\limits_{k=1}^4\psi(y_k),
\end{eqnarray}
\begin{eqnarray}
I_{BB} = \int \prod\limits_{i=1}^4&& d^2x_i d^2y_i
\prod\limits_{k=1}^4\psi(x_k) 
\nonumber\\
&&
\nonumber\\
&&
\times K_{BB}\left[\{x_j\}, \{y_j\}\right]
\prod\limits_{k=1}^4\psi(y_k).
\end{eqnarray}
\label{ints}
\end{mathletters}

\noindent
Equations~(\ref{spea}) can be easily solved and we have for 
the optimal fluctuation 

\begin{mathletters}
\begin{equation}
\bar F_1^{(0)} = \sqrt{\lambda} 
\frac{ I_{FF} + \sigma^2 I_{FB} }
{\sqrt{2\left( I_{FB} + \sigma^2 I_{BB} \right )}},
\end{equation}
\begin{equation}
\bar B^{(0)} = \frac{1}{2} \sqrt{\lambda}
\sqrt{2\left( I_{FB} + \sigma^2 I_{BB} \right )}.
\end{equation}
\label{opt}
\end{mathletters}

\noindent 
Now we only need to find $\lambda$ from the constraint $\sigma^2 =- 
(1+F_0+F_1^{(0)})/B^{(0)}$ or, equivalently, to find
the saddle point solution for $\lambda$. Substituting the saddle point
solutions~(\ref{opt}) into the constraint, we find 

\begin{equation}
\sqrt{\lambda} = -(1+F_0) 
\frac{\sqrt{2\left( I_{FB} + \sigma^2 I_{BB} \right )}}
{I_{FF}+2\sigma^2 I_{FB}+\sigma^4 I_{BB}}.
\label{sl}
\end{equation}

\noindent
The exponent $A_{nl}$ at the saddle point Eq.~(\ref{opt}) is given by

\begin{equation}
\bar A_{nl}(\lambda) = \frac{\lambda}{2\bar B^{(0)}}
\left[ \sigma^2 \bar B^{(0)} + \bar F_1^{(0)} +2(1+F_0) \right].
\label{ac}
\end{equation}

\noindent
Finally, evaluating the exponent~(\ref{ac}) for the optimal value of 
$\lambda$ Eq.~(\ref{sl}), we obtain the probability 
to find an LSD with the spin density amplitude $\sigma$

\begin{equation}
\langle\rho(\sigma^2)\rangle \sim \exp\left(-\frac{(1+F_0)^2}
{2\left(I_{FF}+2\sigma^2 I_{FB}+\sigma^4 I_{BB}\right)}\right),
\label{r2}
\end{equation}
 
\noindent
which for $\sigma = 0$ coincides with the result Eq.~(\ref{r1}) of the 
linear problem, just as the result Eq.~(\ref{r12}) of the qualitative
argument above.

The final step is to rewrite the distribution Eq.~(\ref{r2}) in terms of the
spin $S$ of the LSD. Converting the spin density amplitude 
$\sigma$ in Eq.~(\ref{r2}) to the spin $S$ by means of Eq.~(\ref{spin}), we
obtain the final expression for the spin distribution

\begin{equation}
\rho(S^2) \sim \exp\left ( -\frac{\gamma g^2 (1+F_0)^2}
{1 - 2\alpha \frac{S^2}{g^2} + \beta \frac{S^4}{g^4} }\right),
\label{res}
\end{equation}

\noindent
where the numerical factors are

\begin{mathletters}
\begin{eqnarray}
\alpha = \frac{g^2|I_{FB}|}{I_{FF}\left[\int d^2x \psi(x) \right]^2}
\approx 1.92,
\end{eqnarray}
\begin{eqnarray}
\beta = \frac{g^4I_{BB}}{I_{FF}\left[\int d^2x \psi(x) \right]^4}
\approx 7.36,
\end{eqnarray}
\begin{eqnarray}
\gamma = \frac{1}{2g^2I_{FF}} \approx 3.9.
\end{eqnarray}
\label{fct}
\end{mathletters}

\noindent
The factor $\gamma$ is the same as in the linear problem and is listed here 
for completeness.

The factors Eq.~(\ref{fct}) are now independent of the size $R$ of the LSD.
To see that, one needs to notice that all the correlators $\hat{\cal K}$ depend
on this scale in the same way (when a single LSD is considered, so that $R$ is
the only scale in the integral) $\hat{\cal K}\sim R^{-4}$ [see 
Eq.~(\ref{dkm})]. The wave function $\psi(x)$ is inverse in $R$. Therefore, 
the integrals Eq.~(\ref{ints}), unlike the integral Eq.~(\ref{int}), which 
appears in the solution to the linear problem, do depend on 
$R$, since they contain different number of functions $\psi(x)$

\begin{mathletters}
\begin{equation}
I_{FB} \propto - \frac{R^2}{g^4},
\end{equation}
\begin{equation}
I_{BB} \propto \frac{R^4}{g^6}.
\end{equation}
\label{in}
\end{mathletters}

\noindent
The different size dependence of the integrals Eq.~(\ref{int}) and $I_{FF}$ 
is compensated in Eq.~(\ref{fct}) by additional factors of $\int d^2x \psi(x)$.

The distribution Eq.~(\ref{res}) is thus the same as the qualitative result 
Eq.~(\ref{r11}) only now with coefficients Eq.~(\ref{fct}). The coefficients 
were evaluated numerically using the solution $\psi(x)$, which follows from the
linear problem. The coefficients are positive [the negative 
sign of $I_{FB}$ being taken into account explicitly in Eq.~(\ref{res})]. The 
coefficient $\beta$ is positive, ensuring convergence of the expansion 
Eq.~(\ref{r11}), which again is the way to interpret the distribution 
Eq.~(\ref{res}). In the calculation leading to Eq.~(\ref{res}) the limitation 
to small spins follows from the separation procedure Eq.~(\ref{sep}) since it
is valid only close to the instability, where typical spins are small.

The typical value of spin of the LSD is still given by Eq.~(\ref{r13}). The 
spin $S$ turns out to be large, $1\ll S\ll g$, so on length scales larger
than $R$ or at high enough temperature, LSDs behave as classical moments. 
However, LSDs are extended objects and can have any size with equal 
probability (with the exponential accuracy). Thus their spins can not be 
considered as local moments, especially when discussing their interactions.

\section{Interactions between LSDs}
\label{inter}

The importance of LSDs is that their appearance can dramatically change the
magnetic response of the system. At high enough temperatures, we can consider 
them as independent, classical moments, thus we expect the system to be the 
usual paramagnet with the susceptibility described by the Curie law (but with 
Curie constant different form that of the free electron gas). As the 
temperature becomes smaller, the system might change its ground state in a way
that depends on the interaction between LSDs, in particular on its sign and
typical range.

\begin{figure}
\epsfysize = 4.5cm
\vspace{0.2cm}
\centerline{\epsfbox{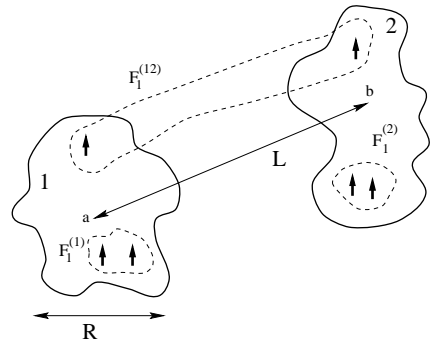}}
\vspace{0.2cm}
\caption{Two LSDs located around points $a$ and $b$ and the three averages
of the non-local susceptibility $F_1$. The two averages $F_1^{(1)}$ and
$F_1^{(2)}$ describe correlations within each single LSD, while $F_1^{(12)}$
describes correlations between different LSDs and thus determines the 
interaction}
\end{figure}

Let us first recall the basic physics of {\it local} moments in a metallic 
system. In a system of local moments there are two competing types of
interaction. First, there is the direct contact interaction \cite{aaa}, which 
sign does not change with distance between the moments, but the amplitude 
decays exponentially beyond the correlation length. This interaction tends to 
turn the system into a ferromagnet. Then there is the RKKY interaction 
\cite{rkk}, which as a function of distance oscillates and decays only as a 
power law ($\sim R^{-2}$ in 2D). The RKKY interaction between local moments 
tends to form a spin glass at low temperatures \cite{spz}. 

Our case is different.
As we have shown in this paper, in a metal close to Stoner instability there
is a non zero probability for LSDs to be spontaneously formed. This 
probability is independent of the size of LSDs, so that droplets of all sizes
can appear. Therefore interaction between the LSDs can not be described in the 
same way as interaction between local moments. Rather, it is given by the 
{\it non-local} susceptibility $F_1(x_1, x_2)$, averaged over the area of the 
interacting LSDs. Calculation of the average, previously denoted as 
$F_1^{(0)}$, involves integration of $F_1(x_1, x_2)$ with two functions
$\psi(x-a)$ which describe the spatial profile of the LSDs. In this section we
use the same Gaussian functions we used to evaluate the integrals 
Eq.~(\ref{int}) and Eq.~(\ref{ints}). Only now these functions carry 
explicitly the dependence on the coordinate $a$ of the center of the LSD.
 
Consider now two LSDs separated by distance $L$ much larger than the size of
both LSDs $L\gg R$, see Fig. 1. In this argument we take both LSDs to be of 
the same size
$R$, but it can be easily generalized for the case where the interacting LSDs
differ is size substantially. Choosing different combinations of the 
wavefunctions $\psi(x-a)$ and $\psi(y-b)$ describing the two LSDs,
we can form three different averages

\begin{mathletters}
\begin{eqnarray}
&&F_1^{(1)} = 
\int d^2x_1 d^2x_2F_1(x_1, x_2)\psi(x_1-a)\psi(x_2-a), \\
&&F_1^{(2)} = 
\int d^2y_1 d^2y_2F_1(y_1, y_2)\psi(y_1-b)\psi(y_2-b), \\
&&F_1^{(12)} = 
\int d^2x d^2yF_1(x, y)\psi(x-a)\psi(y-b). 
\end{eqnarray}
\label{ave}
\end{mathletters}

\noindent
These averages, as random quantities, have a Gaussian distribution 
with the weight determined by the correlation function
$\langle F_1 F_1 \rangle$. We now assume that the LSDs have been already 
formed. The saddle point solution $\bar F_1$ [see Eq.~(\ref{ofl})] describes 
a single LSD, while the interactions are determined by small deviations from 
the saddle point. Therefore the weight of the distribution of the averages 
Eq.~(\ref{ave}) can be evaluated at the ``optimal fluctuation'' point 
Eq.~(\ref{ofl}). Then the distribution can be written as

\begin{eqnarray}
w[&&F_1^{(1)}, F_1^{(2)}, F_1^{(12)}] \propto \nonumber\\
&&
\nonumber\\
&&
\exp\left[ -\int 
\pmatrix{
F_1^{(1)} & F_1^{(2)} & F_1^{(12)}  \cr
 }
\hat{\cal L}^{-1}
\pmatrix{
F_1^{(1)} \cr
F_1^{(2)} \cr 
F_1^{(12)}
}
\right],
\label{dis}
\end{eqnarray}

\noindent
with the weight matrix

\begin{eqnarray}
\hat{\cal L} = 
\pmatrix{
I_{FF} & J_1    & J_2 \cr
J_1    & I_{FF} & J_2 \cr
J_2    & J_2    & J_3 
}.
\label{l}
\end{eqnarray}

\noindent 
The elements of the weight matrix $\hat{\cal L}$ are obtained by averaging 
the correlation function $K_{FF}$ over the area of the LSDs,
i.e. integrating with four wavefunctions $\psi(x-a)$, similar to the integral
$I_{FF}$ [see Eq.~(\ref{int})]. The difference from the case of the single LSD
is that now all but two of the elements of $\hat{\cal L}$ depend on two
different lengthes : the size of the LSD $R$ and the distance $L$ between 
them. Thus the estimate Eq.~(\ref{dkm}) for the correlation function $K_{FF}$
does not apply. The elements $J_i$ are given by the integrals

\begin{mathletters}
\begin{eqnarray}
J_1 = &&\int d^2x_1 d^2x_2 d^2y_1 d^2y_2 
K_{FF}\left[\{x_j\}, \{y_j\}\right]\psi(x_1-a) 
\nonumber\\
&&
\nonumber\\
&&
\times\psi(x_2-a)
\psi(y_1-b) \psi(y_2-b) \approx \frac{0.15}{\pi^2 g^2} \frac{R^4}{L^4} ,
\end{eqnarray}
\begin{eqnarray}
J_2 = \int  &&  d^2x_1d^2x_2 d^2y_1 d^2y_2 
K_{FF}\left[\{x_j\}, \{y_j\}\right]
\nonumber\\
&&
\nonumber\\
&&
\times\psi(x_1-a) \psi(x_2-b)
\psi(y_1-b) \psi(y_2-b) 
\nonumber\\
&&
\nonumber\\
&&
\approx \frac{1}{6\pi^2 g^2} \frac{R^4}{L^4} 
\ln^2\frac{R}{L},
\end{eqnarray}
\begin{eqnarray}
J_3 = \int  &&  d^2x_1d^2x_2 d^2y_1 d^2y_2 
K_{FF}\left[\{x_j\}, \{y_j\}\right]
\nonumber\\
&&
\nonumber\\
&&
\times\psi(x_1-a) \psi(x_2-b)
\psi(y_1-a) \psi(y_2-b) 
\nonumber\\
&&
\nonumber\\
&&
\approx \frac{1}{6\pi^2 g^2} \frac{R^4}{L^4} 
\ln^2\frac{R}{L}.
\end{eqnarray}
\label{in2}
\end{mathletters}

\noindent 
The integrals Eq.~(\ref{in2}) were calculated in the leading order in $R/L$.
Clearly, $J_2 \ne J_3$ exactly, but the difference comes in the numerical 
factor under the logarithm, which we here neglect. This does not have any 
bearing on our conclusions. The $1/L^4$ dependence of all the integrals 
follows from the frequency integral in Eq.~(\ref{cff}), which is determined
by the Thouless energy corresponding to the largest length in the problem,
which is now $L$. 

The integral $I_{FF}$ is independent of all length scales and therefore is 
much larger than any of $J_i$. Thus the weight matrix $\hat{\cal L}^{-1}$ in 
Eq.~(\ref{dis}) to the leading order in $R/L$ is 

\begin{eqnarray}
\hat{\cal L}^{-1} \approx 0.8 \pi^2 g^2
\pmatrix{
1   & 0   & -1 \cr
0   & 1   & -1 \cr
-1  & -1  & f^{-1} 
},
\label{L}
\end{eqnarray}

\noindent 
where the dimensionless function $f(R/L)$ is given by

\begin{equation}
f(R/L) \approx 0.72 \frac{R^4}{L^4} \ln^2 \frac{R}{L}.
\label{c}
\end{equation}

To describe the interaction between LSDs we need to find the distribution 
of $F_1^{(12)}$ under the condition that the two LSDs exist, namely that
$F_1^{(1)}<0$ and $F_1^{(2)}<0$. This is given by the {\it conditional}
probability distribution

\begin{mathletters}
\begin{eqnarray}
W[F_1^{(12)}]&&  =\frac{1}{N^2}
\int {\cal D}[F_1^{(1)}, F_1^{(2)}] 
\theta(-1-F_0-F_1^{(1)})
\nonumber\\
&&
\nonumber\\
&&
\times\theta(-1-F_0- F_1^{(2)}) w[F_1^{(1)}, F_1^{(2)}, F_1^{(12)}] ,
\end{eqnarray}
\begin{eqnarray}
N = \int {\cal D}[F_1^{(1)}]e^{-{{(F_1^{(1)})^2}\over{I_{FF}}}}
\theta(-1-F_0- F_1^{(1)}),
\end{eqnarray}
\label{w}
\end{mathletters}

\noindent
where $w$ is the distribution Eq.~(\ref{dis}). Since the weight matrix in
Eq.~(\ref{dis}) is a $c$-number, the integrals in Eq.~(\ref{w}) are usual
Gaussian integrals and not functional integrals. The $\theta$-functions in 
Eq.~(\ref{w}) make the exact integration in terms of elementary
functions impossible, but we can use the small parameter $f(R/L)\ll 1$ to 
estimate $W[F_1^{(12)}]$ with exponential accuracy, which is all we need
to describe interaction between LSDs. Up to the pre-exponential factor

\begin{equation}
W[F_1^{(12)}] \propto 
\exp\left[-\frac{0.8\pi^2 g^2}{f}\left(F_1^{(12)}
+2(1+F_0)f\right)^2\right],
\label{re}
\end{equation}

\noindent
so that the distribution is a Gaussian (as it should be since we considered
$F_1$ to be a Gaussian random quantity from the very beginning).

The average $F_1^{(12)}$ given by the distribution Eq.~(\ref{re}) is shifted
from zero to the negative value $-2f(R/L)(1+F_0)$, i.e. the average 
interaction appears to be ferromagnetic. However, the distribution
Eq.~(\ref{re}) also allows for strong fluctuations of $F_1^{(12)}$. These
fluctuations can be estimated as

\begin{equation}
\frac{\langle (\delta F_1^{(12)})^2 \rangle}{\langle F_1^{(12)} \rangle^2}
\sim \frac{1}{g^2(1+F_0)^2 f}.
\end{equation}

\noindent
The function $f$ [see Eq.~(\ref{c})] decreases with the distance between LSDs.
Therefore at large enough distances the fluctuations of $F_1^{(12)}$ exceed the
average and the sign of $F_1^{(12)}$ becomes random. 

The cross-over distance $L^*$ can be estimated as (from the condition 
$\langle (\delta F_1^{(12)})^2 \rangle/\langle F_1^{(12)} \rangle^2 \sim 1$)

\begin{equation}
L^* \sim R \sqrt{g(1+F_0)} \gg R.
\label{L*}
\end{equation}

\noindent
This distance has to be compared with the typical distance between LSDs. The
latter can be estimated as follows. The concentration of LSDs has to be 
proportional to the probability Eq.~(\ref{r1})

\begin{equation}
n \propto {1\over{R^2}} e^{-\gamma g^2(1+F_0)^2},
\label{n}
\end{equation}

\noindent
where $R$ is the characteristic length of a LSD. Therefore the typical
distance between LSDs is exponentially large in the parameter $g(1+F_0)\gg 1$,
while the cross-over length $L^*$ is large only as a power of the same
parameter. Thus, the interaction between typical LSDs has random sign.

The energy of interaction between typical LSDs decays as the second power of 
the distance between them

\begin{equation}
U_t = {1\over{\nu g L^2}} \sim {1\over{\nu g R^2}} e^{-\gamma g^2(1+F_0)^2},
\label{u}
\end{equation}

\noindent
where $\nu$ is the density of states and we have estimated the typical 
distance between LSDs from Eq.~(\ref{n}).

The results of this Section resemble the results of Ref.~\onlinecite{spz},
where the electron-mediated interaction between magnetic moments in two 
ferromagnets separated by a disordered metal was also found to have a random 
sign. The diference between our model and that of Ref.~\onlinecite{spz} is 
that in our case the ferromagnetic regions (LSDs) are created by the very same 
impurity configurations as those responsible for the interaction between LSDs. 
Therefore, the interaction between LSDs can be considered only under the 
condition of their existance and thus is characterized by the 
{\it conditional} probability distribution Eq.~(\ref{re}). As a result, the 
average $F_1^{(12)}$ is negative, i.e. ferromagnetic, and dominates the
fluctuations at distances smaller than $L^*$. However, since the typical
distance between LSDs is exponentially large, the typical interaction has
random sign.  

This fact is not surprising, since we are dealing with a disordered system
where one could expect to find some spin glass phase (at $T=0$, similar to the 
case of a superconductor in weak magnetic field, considered in 
Ref.~\onlinecite{spi}. In the model considered in Ref.~\onlinecite{spi} 
mesoscopic fluctuations become uncorrelated beyond the magnetic length. In our 
case the role similar to that of the magnetic field in Ref.~\onlinecite{spi} 
would have been played by the spin-orbit coupling, which we have so far 
neglected. However, taking the spin-orbit coupling into account does not 
change the main results of this paper. As we have shown above, the typical 
interaction between LSDs is random due to large fluctuations in Eq.~(\ref{re}).
Should we include the spin-orbit coupling, we would need to compare the 
spin-orbit length to $L^*$ in order to determine the length beyond which the 
interaction becomes random. But since the typical distance between LSDs is 
exponentially larger than $L^*$ the exact value of such cross-over length is 
not very important.

\section{contribution to physical quantities}
\label{phys}

In this section we estimate the contribution of the LSDs to observable
physical quantities. We consider the examples of paramagnetic susceptibility
and the dephasing time. To be observable, the contribution of LSDs should
exceed the regular contribution of the electron system. We show that, for 
both quantities, it happens at the same temperature $T^*$, below which the
temperature dependence of both quantities changes. The Pauli susceptibility
crosses over to a Curie-type $1/T$ dependence, while the dephasing time
saturates and becomes temperature independent. 

Since the typical interaction energy Eq.~(\ref{u}) is exponentially small,
LSDs behave as weakly interacting moments and their 
contribution to the paramagnetic susceptibility is given by the
Curie law $\chi_{LSD} = C/T$, where the Curie constant $C$ is proportional
to the square of the spin $S$ of the LSD [see Eq.~(\ref{r13})] and the
density (or concentration) of LSDs Eq.~(\ref{n}).  We have calculated the 
probability Eq.~(\ref{r1}) with the exponential accuracy, therefore the 
pre-exponential factor in the density of LSDs should be given by the typical 
length characterizing the distribution of LSDs. This dependence could be 
elucidated from the following dimensional argument. The probability to find 
an LSD in unit volume (or, rather, area, which we still denote by $V$) of size 
in between $R$ and $R+dR$, with spin in between $S$ and $S+dS$ is given by the 
distribution 

\begin{equation}
d{\cal W} = \rho(S^2) dV dS d^2R
\end{equation}

\noindent
Clearly, the dimensionality of $\rho(S^2)$ is $L^{-4}$, where $L$ is the
characteristic length. Since the exponential in $\rho(S^2)$ 
[see Eq.~(\ref{res})] is independent of length, the size of a LSD can not
be pinned to any length in the system. Since individual LSDs contribute to
Curie susceptibility independently, we need to sum over all possible sizes
and given the $L^{-4}$ dependence, the main contribution comes from LSDs 
of the smallest possible size, namely the mean-free path $l$. Thus, 
contribution to the susceptibility (up to a numerical coefficient) is

\begin{equation}
\chi_{LSD} \propto {1\over{\tau T}} \frac{\nu}{g(1+F_0)^2} 
e^{-\gamma g^2(1+F_0)^2},
\label{ch}
\end{equation}

\noindent
where $\tau$ is the scattering time. Here we used the typical 
value~(\ref{spin}) of the spin of a LSD. 
This contribution has to be compared 
with the Pauli susceptibility Eq.~(\ref{pauli}), which is temperature 
independent. As a result, the contribution Eq.~(\ref{ch}) dominates at
temperatures smaller than $T^*$, which up to numerical factors is

\begin{equation}
T^* \sim  {1\over{\tau}} \frac{1}{g(1+F_0)} 
e^{-\gamma g^2(1+F_0)^2}.
\label{tstar}
\end{equation}

\noindent
This temperature has to be compared with the typical interaction energy 
Eq.~(\ref{u}). Substituting the mean-free path $l$ for the chracteristic
size $R$ in Eq.~(\ref{u}) we find the ratio $U_t / T^*$ to be 

\begin{equation}
{{U_t}\over{T^*}} \sim \frac{(1+F_0)^2}{g(1+F_0)} \ll 1.
\label{ratio}
\end{equation}

\noindent
Therefore, the Curie behavior Eq.~(\ref{ch}) persists over a wide temperature 
range $U_t < T < T^*$.

Similarly, we estimate the contribution of LSDs to the dephasing time
$\tau_{\varphi}$. LSDs are regions where the impurity configuration makes
it energetically favorable for the electrons to align their spins. Some
other electron entering such region will ``feel'' the overall polarization
as if it were magnetic field. The corresponding dephasing time can be 
estimated as the square of the Zeeman energy divided by the Thouless energy
(at the size of the LSD). More formally, since the interaction with the 
polarization stems from the exchange interaction, we can find $\tau_{\varphi}$
from the perturbation theory \cite{alt}

\begin{eqnarray*}
{1\over{\tau_{\varphi}}} = {F_0^2\over{\nu^2}} \int \frac{d^2 r_1}{V}
\int \frac{d^2 r_2}{V} \langle \sigma(r_1) \sigma(r_2)\rangle 
{\cal D}(r_1-r_2),
\end{eqnarray*}

\noindent
where ${\cal D}(r_1-r_2)$ is the (electron) diffusion propagator and
$\langle \sigma(r_1) \sigma(r_2)\rangle$ is the correlation function of the
spins of LSDs. We can now estimate $\tau_{\varphi}$ as

\begin{equation}
{1\over{\tau_{\varphi}}} \sim {F_0^2\over{\nu^2}} \frac{n S^2}{D}
\sim {F_0^2\over{\tau}} \frac{1}{g^2(1+F_0)^2} 
e^{-\gamma g^2(1+F_0)^2},
\label{t}
\end{equation}

\noindent
where again the dominant contribution comes from LSDs of smallest size 
[$n$ is the concentration of LSDs, see Eq.~(\ref{n})]. This should be compared 
with the contribution of the usual Gaussian spin fluctuations \cite{us}

\begin{equation}
{1\over{\tau_{\varphi}^{s}}} \sim {2F_0^2\over{(1+F_0)(2+F_0)}} 
\frac{T}{g} \ln g(1+F_0).
\end{equation}

\noindent
Again, up to the numerical factors, the contribution of LSDs dominates 
at temperatures lower than $T^*$ [given by Eq.~(\ref{tstar})]. 

The actual crossover temperatures for different physical quantities might 
differ by a factor of order $\ln g(1+F_0)$, but such difference is beyond
the accuracy of our treatment. However, the discussed contribution of LSDs
to physical quantities suggests that our scenario of the magnetic 
fluctuations in a metal close to the Stoner instability can be experimentally
observed. The LSDs lead to the saturation of the dephasing time at low 
temperatures. If such saturation is observed, one should look at the behavior 
of the paramagnetic susceptibility in the same temperature range. If the LSDs
are present in the system, then the onset of the Curie-like temperature
dependence should also be detected.

\section{conclusions}

In this paper we have considered the effect of disorder on magnetic properties
of the ground state of a metal close to the Stoner instability. We have shown
that even though on the mean field level the ground state of the metal is 
paramagnetic ($1+F_0 > 0$), there is non zero (exponentially small) probability
to form local spin droplets, i.e. domains of non zero spin polarization.
The probability to form a LSD is independent of its size $R$, thus LSDs of any 
size can appear. The total spin of the LSD is also independent of its size 
and obeys the distribution Eq.~(\ref{r12}), with the typical value 
$S\simeq (1+F_0)^{-1}$, which is large $1\ll S \ll g$, so that LSDs can be 
considered as classical spins. 

Considered as independent moments, LSDs contribute to the observables, 
changing the temperature dependence of both the paramagnetic susceptibility
and the dephasing time at temperatures lower than certain cross-over
temperature Eq.~(\ref{tstar}). When $T<T^*$, the dephasing time saturates
to the temperature-independent value Eq.~(\ref{t}), while the susceptibility
acquires the Curie-like $1/T$ dependence.

Both the Curie susceptibility Eq.~(\ref{ch}) and the dephasing time 
Eq.~(\ref{t}) were obtained in approximation of non-interacting LSDs. This
approach is valid at temperatures larger than the typical value~(\ref{u}) of
the interaction between LSDs. Since the cross-over temperature $T^*$ is much
larger than the interaction~(\ref{u}), there is a parametrically wide 
temperature regime [by our large parameter $g(1+F_0)$, see Eq.~(\ref{ratio})],
$U_t < T < T^*$, where the Curie behavior of the susceptibility and the 
saturation of the dephasing time can be observed. At smaller temperatures
$T<U_t$, however, the LSDs can not be considered as non-interacting moments
and the behavior of the system changes. Interaction of LSDs with each other 
or with itinerant electrons should lead either to screening of the local spins
or to forming some spin glass state (due to the random sign of the 
interaction). Such regime was not considered in this paper..

We acknowledge helpful discussions with L.B. Ioffe and B. Spivak. We also 
acknowledge the hospitality of Theoretische Physik III, Ruhr-Universit\"at 
Bochum, where this work was started.
I.A. is Packard research fellow. A.L. is supported by the NSF under Grant
9812340.

\appendix

\section*{}

Here we discuss the correlation functions that describe the mesoscopic 
fluctuations in the system, i.e. give the weight in the distribution  
Eq.~(\ref{corr}) of the random quantities $F_1$ and $B$.
Both quantities are the coefficients in the expansion of the thermodynamic
potential Eq.~(\ref{tpnl}) in powers of the spin density. Therefore the 
diagrams for $F_1$ and $B$ can be obtained by differentiating the 
diagram for the exact thermodynamic potential. To calculate the 
correlators one has to multiply the random quantities and then average 
over the disorder. As a result we get three different correlators, 
depicted diagrammatically on Figs. 2 - 4.

In terms of exact electronic Green's functions the thermodynamic potential
can be written as

\begin{equation}
\Omega = \int \frac{d\epsilon}{2\pi i} f(\epsilon) {\bf Tr} \ln 
\frac{G^R(\epsilon)}{G^A(\epsilon)},
\label{o}
\end{equation}

\noindent
where the Green's function is defined in the fluctuating field 
$\vec\sigma$, i.e.

\begin{equation}
G^{R(A)}(\epsilon) = 
\left[\epsilon -\hat{\cal H} - \vec\tau\vec\sigma\pm i0 \right]^{-1}.
\end{equation}

\noindent
Here $\hat{\cal H}$ is the Hamiltonian of the system, $\tau^j$ are the Pauli
matrices. The symbol ${\bf Tr}$ denotes the trace over spin indices and the 
integration over all spatial cooridinates. For brevity we do not explicitly
indicate that $G^{R(A)}(\epsilon)$ depend on spatial coordinates (due to
disorder in $\hat{\cal H}$).

The expansion~(\ref{tp}) can be achieved by taking variational derivatives 
of the thermodynamic potential~(\ref{o}) with respect to $\vec\sigma$. 
The second 
variational derivative of the thermodynamic potential corresponds to the 
second order term in Eq.~(\ref{tp}),

\begin{eqnarray}
\frac{\delta^2\Omega}{\delta\sigma^\alpha(x_1)\delta\sigma^\beta(x_2)}&&
= \int \frac{d\epsilon}{2\pi i} f(\epsilon)
\nonumber\\
&&
\nonumber\\
&&
\times{\bf Tr'} \Big [G^R \tau^\alpha G^R \tau^\beta -
G^A \tau^\alpha G^A \tau^\beta\Big].
\label{sed}
\end{eqnarray}

\noindent
Here the prime in ${\bf Tr'}$ means that there is no integration over the 
coordinates in the left-hand side (in this case $x_1$
and $x_2$). The trace over the Pauli matrices gives (in the absence of 
spin-orbit coupling) ${\rm Tr}[\tau^\alpha\tau^\beta] = 2\delta^{\alpha\beta}$.
Upon subtracting the average, the derivative~(\ref{sed}) is proportional to 
the non-local quantity $F_1(x_1, (x_2)$:

\begin{eqnarray}
F_1(x_1, x_2) = &&
\frac{\delta^2\Omega}{\delta\sigma^\alpha(x_1)\delta\sigma^\beta(x_2)}
\nonumber\\
&&
\nonumber\\
&&{\; \; \; \; \; \; \; \; \; \; \; \; }
- \langle\frac{\delta^2\Omega}
{\delta\sigma^\alpha(x_1)\delta\sigma^\beta(x_2)}\rangle.
\label{f1}
\end{eqnarray}

\noindent
Similarly, the coefficient $B$ in Eq.~(\ref{tp}) is given by the fourth
variational
derivative of the thermodynamic potential~(\ref{o}) (only the average is zero
in this case):

\begin{eqnarray}
B[\{x_i\}] = \int \frac{d\epsilon}{2\pi i} && f(\epsilon) {\bf Tr'} 
\Big [G^R \tau^\alpha G^R \tau^\beta
G^R \tau^\mu G^R \tau^\nu 
\nonumber\\
&&
\nonumber\\
&&
- G^A \tau^\alpha G^A \tau^\beta 
G^A \tau^\mu G^A \tau^\nu\Big ].
\label{fd}
\end{eqnarray}

\noindent
The trace over the Pauli matrices has two parts, 
${\rm Tr}[\tau^\alpha\tau^\beta\tau^\mu\tau^\nu] =
2 [\delta^{\alpha\beta}\delta^{\mu\nu} - 
(\delta^{\alpha\mu}\delta^{\beta\nu} - \delta^{\alpha\nu}\delta^{\beta\mu})]$.
The first part corresponds to the fourth order term in Eq.~(\ref{tp}), while 
the second part is the cross-product term, which disappears since we consider
only the singlet fluctuations $\vec\sigma = (0, 0, \sigma)$.  

We are now ready to calculate correlation functions of the mesoscopic
fluctuations $F_1$ and $B$. Averaging over disorder is performed in a standard
manner (see Ref.~\onlinecite{alt} for details). The correlator of the 
fluctuation $F_1(x_1, x_2)$

\begin{eqnarray}
K_{FF}&&[\{x_j\}, \{y_j\}] = 
\nonumber\\
&&
\nonumber\\
&&
K_{FF}(x_1, x_2, y_1, y_2) = 
\langle F_1(x_1, x_2) F_1(y_1, y_2) \rangle
\label{ff}
\end{eqnarray}

\noindent
is given by a sum of four diagrams, one of which is depicted on Fig. 2.

\begin{figure}
\epsfysize = 7cm
\vspace{0.2cm}
\centerline{\epsfbox{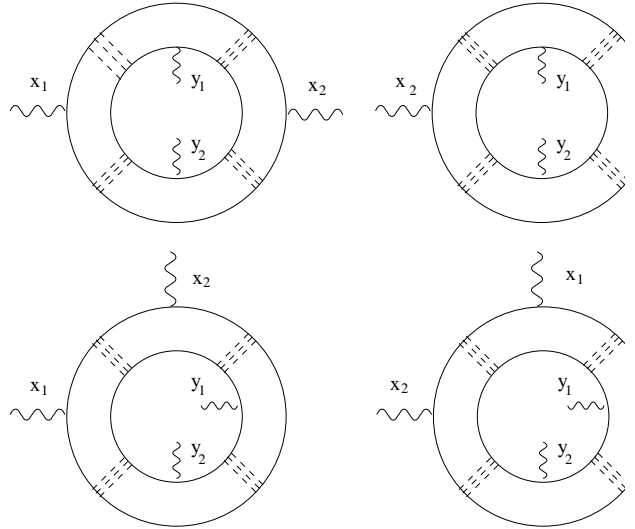}}
\vspace{0.2cm}
\caption{Diagrams for the averaged correlation function of the 
non-linear part of the paramagnetic susceptibility.}
\end{figure}

\noindent
All four diagrams contribute equally to the integral Eq.~(\ref{int}) (as
will be clear from the explicit form of $K_{FF}$), 
therefore we shall proceed evaluating the contribution of the diagrams 
on Fig. 2. After averaging, each diagram 
corresponds to a product of four diffusion propagators (or diffusons) and 
four vertex blocks. 

The diffuson in momentum representation is given by

\begin{eqnarray*}
{\cal D}(\omega; q) = \frac{1}{-i\omega+Dq^2},
\end{eqnarray*}

\noindent
where $D$ is the diffusion constant, and $q$ is a 2D momentum 
vector. Each vertex block (see Fig. 4) contains precisely the same
factors $2\pi\nu\tau^2$ as each diffuson, thus in momentum representation
the diagrams on Fig. 2 are expressed in terms of the integral

\begin{eqnarray}
K_{FF}&&[\{k_j\}] = 
\int\limits^0_{-\infty} \frac{d\epsilon_1}{\pi\nu}
\int\limits^0_{-\infty} \frac{d\epsilon_2}{\pi\nu}
\int d^2 Q \; {\cal D}(\omega; Q)
\nonumber\\
&&
\nonumber\\
&&
\times{\cal D}(\omega; Q-k_1) \; {\cal D}(\omega; Q-k_3) \; 
{\cal D}(\omega; Q-k_1-k_4)
\nonumber\\
&&
\nonumber\\
&&
{\; \; \; \; \; \; \; \; \; }
\times\delta(k_1+k_4-k_2-k_3),
\label{kffm}
\end{eqnarray}

\noindent
where $\omega=\epsilon_1-\epsilon_2$.

For the purposes of this 
paper it is more convenient to write the correlator Eq.~(\ref{ff})
in the coordinate representation as (we have also evaluated one of the
frequency integrals)

\begin{eqnarray}
K_{FF}&&[\{x_j\}, \{y_j\}] = 
- \int\limits_{-\infty}^\infty \frac{d\omega}{2\pi^2\nu^2} \; |\omega| \; 
{\cal D}(\omega; y_1 - x_1)
\nonumber\\
&&
\nonumber\\
&&
\times{\cal D}(\omega; x_1 - y_2) \;
{\cal D}(\omega; x_2 - y_1) \;
{\cal D}(\omega; y_2 - x_2),
\label{cir}
\end{eqnarray}

\noindent
where $x_i$ and $y_i$ are 2D coordinate vectors and

\begin{equation}
{\cal D}(\omega; x) = 
\int\frac{d^2 q}{(2\pi)^2} 
{\cal D}(\omega; q) e^{-i q x}
\label{ft}
\end{equation}

\noindent
is the Fourier transform of the diffusion denominator to the position
space.

The remaining frequency integral can be evaluated by using the following
integral representation for ${\cal D}(\omega; x)$. First, we represent
the diffusion denominator as an integral over an auxiliary variable 

\begin{eqnarray*}
{1\over{-i\omega + Dq^2}} = \int\limits_0^\infty dt 
\exp\big[-t(-i\omega + Dq^2)\big].
\end{eqnarray*}

\noindent
The momentum integral in Eq.~(\ref{ft}) becomes Gaussian and we obtain

\begin{eqnarray}
{\cal D}(\omega; x) = \frac{1}{4\pi D} \int\limits_0^\infty 
\frac{dt}{t} \; \exp\left[i\omega t - \frac{x^2}{4tD}\right].
\label{ir}
\end{eqnarray}

\noindent
Substituting the integral representation Eq.~(\ref{ir}) into the 
correlator Eq.~(\ref{cir}), we find the final expression for the 
contribution of the diagram on Fig. 2 to the correlator Eq.~(\ref{ff})

\begin{eqnarray}
&&K_{FF}[\{x_j\}, \{y_j\}] = \frac{1}{(2\pi\nu D)^2}\frac{1}{4\pi^4}
\int\limits_0^\infty \frac{dt_1 dt_2 dt_3 dt_4}{t_1 t_2 t_3 t_4}
\nonumber\\
&&
\nonumber\\
&&
\frac{\exp\left[ - \frac{(x_1-y_1)^2}{t_1} - \frac{(x_2-y_1)^2}{t_2}
 - \frac{(x_1-y_2)^2}{t_3} - \frac{(x_2-y_2)^2}{t_4}\right]}
{(t_1+ t_2+ t_3+ t_4)^2}
\label{cff}
\end{eqnarray}

To estimate the size of the fluctuation regions with non zero spin (LSD)
we need the explicit dependence of the correlator Eq.~(\ref{cff}) on the 
parameters of the problem. To do that we introduce a length scale $R$
which characterizes the size if the LSD and write 
Eq.~(\ref{cff}) as

\begin{eqnarray}
K_{FF}&&[\{x_j\}, \{y_j\}] = \frac{1}{g^2R^4} \tilde K_{FF}
\label{dk}
\end{eqnarray}

\noindent
where $\tilde K_{FF}$ is the dimensionless counterpart of the integral
Eq.~(\ref{cff}). 

\begin{figure}
\epsfysize = 4cm
\vspace{0.2cm}
\centerline{\epsfbox{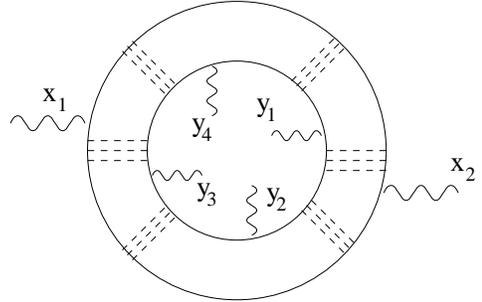}}
\vspace{0.2cm}
\caption{Typical diagram for the averaged correlation function of 
$\protect\langle F_1 B \protect\rangle$. The rest of the diagrams 
are obtained by shifting positions of the spin vertices on the inside line
relative to those on the outside line and interchanging coordinate indices
similar to the diagrams on Fig. 2. There are 3 topologically different 
diagrams.}
\end{figure}

\begin{figure}
\epsfysize = 5.5cm
\vspace{0.2cm}
\centerline{\epsfbox{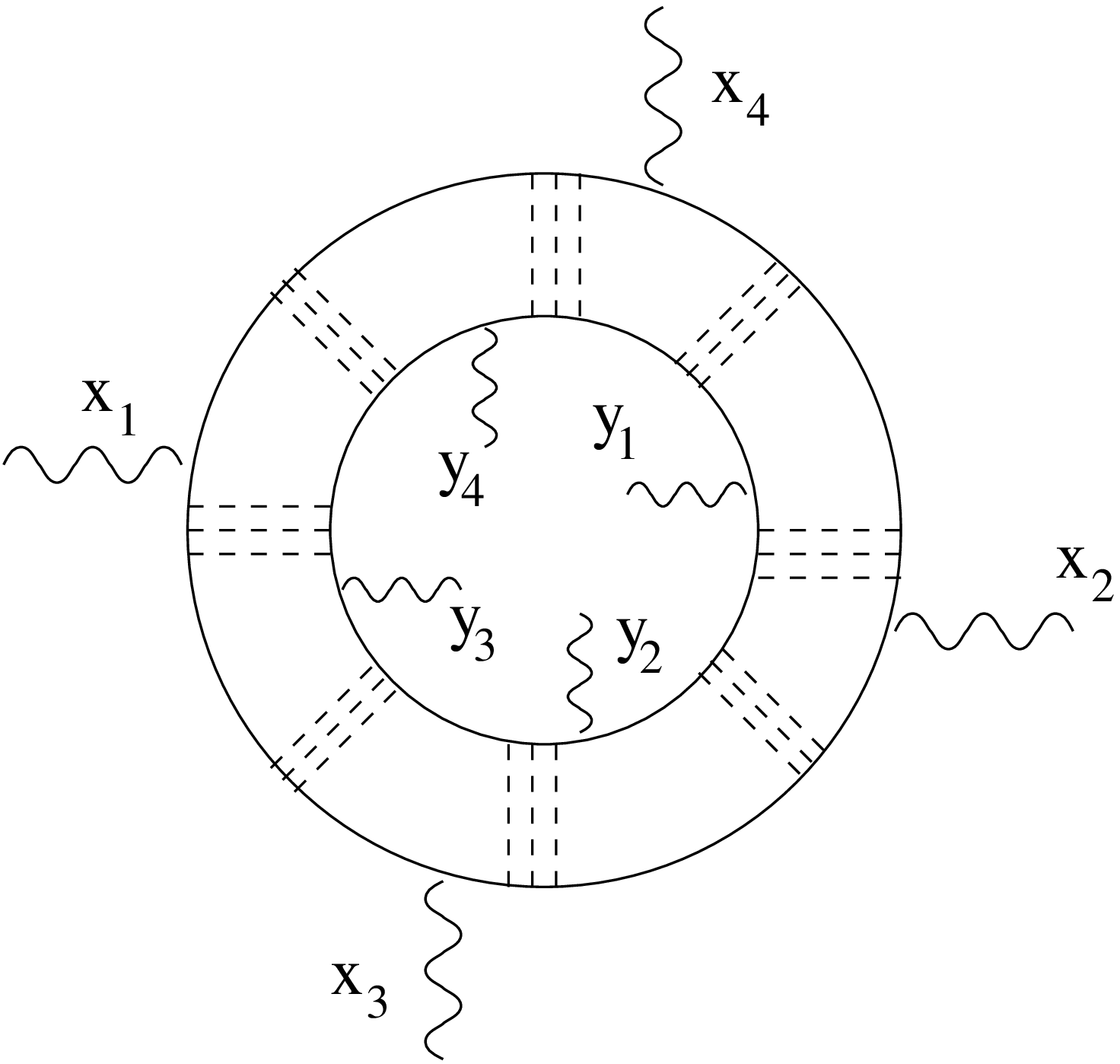}}
\vspace{0.2cm}
\caption{Typical diagram for the averaged correlation function of 
$\protect\langle B B \protect\rangle$. The rest of the diagrams 
are obtained by shifting positions of the spin vertices on the inside line
relative to those on the outside line and interchanging coordinate indices
similar to the diagrams on Fig. 2.
There are 8 tpologically different diagrams.}
\end{figure}

The remaining correlators in Eq.~(\ref{corr}) are constructed in the same 
manner as Eq.~(\ref{cff}), the only difference being the number of 
diffusons. The corresponding diagrams are given in Fig. 3 and 4. 
Again, the total number of diagrams is large, therefore we give expressions 
for the typical contributions depicted on Figs. 3 and 4. The rest of the 
diagrams are obtained by interchanging coordinate indices.

The diagram on Fig. 3 contains six diffusons and six vertex blocks. Since
each of the vertex blocks carries a factor of $i$, the overall
sign of the diagram is negative. Using the integral representation 
Eq.~(\ref{ir}) for diffusion denominators, we find 

\begin{eqnarray}
K_{FB}[\{x_j\}, \{y_j\}] = - &&\frac{1}{2(2\pi)^4}\frac{1}{(2\pi D)^4}
\nonumber\\
&&
\nonumber\\
&&
\times\int\limits_0^\infty \prod\limits_{k=1}^6\frac{dt_k}{t_k}
\frac{\exp [-T_{FB}]}
{(\sum\limits_{i=1}^6 t_i)^2}
\label{cfb}
\end{eqnarray}
\begin{eqnarray*}
T_{FB} = &&\frac{(x_1-y_3)^2}{t_1} + \frac{(x_1-y_4)^2}{t_2}
 + \frac{(x_2-y_2)^2}{t_3}
\nonumber\\
&&
\nonumber\\
&&
{\; \; \; \; \; \; \; \; }
 + \frac{(x_2-y_1)^2}{t_4}
 + \frac{(y_1-y_4)^2}{t_5} + \frac{(y_3-y_2)^2}{t_6}
\end{eqnarray*}

The correlator on Fig. 4 can be written in the same way. It has positive sign, 
since it contains eight vertex blocs.

\begin{eqnarray}
K_{BB}[\{x_j\}, \{y_j\}] = &&\frac{1}{8(2\pi)^4}\frac{1}{(2\pi D)^6}
\nonumber\\
&&
\nonumber\\
&&
\times\int\limits_0^\infty \prod\limits_{k=1}^8\frac{dt_k}{t_k}
\frac{\exp [-T_{BB}]}
{(\sum\limits_{i=1}^8 t_i)^2}
\label{cbb}
\end{eqnarray}
\begin{eqnarray*}
T_{BB} = &&\frac{(x_1-y_3)^2}{t_1} + \frac{(x_1-y_4)^2}{t_2}
 + \frac{(x_2-y_2)^2}{t_3}
\nonumber\\
&&
\nonumber\\
&&
{\; \; \; \; \; \; \; \; }
 + \frac{(x_2-y_1)^2}{t_4}
 + \frac{(x_3-y_3)^2}{t_5} + \frac{(x_3-y_2)^2}{t_6}
\nonumber\\
&&
\nonumber\\
&&
{\; \; \; \; \; \; \; \; }
{\; \; \; \; \; \; \; \; }
 + \frac{(x_4-y_1)^2}{t_7} + \frac{(x_4-y_4)^2}{t_8}
\end{eqnarray*}

Altogether, these correlators can be combined in the matrix Eq.~(\ref{corr})

\begin{eqnarray}
\hat{\cal K} = 
\pmatrix{
K_{FF}[\{x_j\}, \{y_j\}] & K_{FB}[\{x_j\}, \{y_j\}] \cr
K_{FB}[\{x_j\}, \{y_j\}] & K_{BB}[\{x_j\}, \{y_j\}] \cr
 }.
\label{cr}
\end{eqnarray}

\noindent
Similarly to Eq.~(\ref{dk}), the dimensional analysis gives the size 
dependence in terms of the scale $R$

\begin{eqnarray}
\hat{\cal K} = \frac{1}{g^2R^4}
\pmatrix{
\tilde K_{FF}               & -\frac{1}{g^2}\tilde K_{FB} \cr
-\frac{1}{g^2}\tilde K_{FB} &  \frac{1}{g^4}\tilde K_{BB} \cr
 }.
\label{dkm}
\end{eqnarray}

\noindent
The sign of the correlator follows from the diagrams, so that the 
dimensionless integrals $\tilde K$ are positive numbers.

\end{multicols}

\end{document}